\documentclass[9pt,twocolumn,twoside]{osajnl}
\journal{ol} % Choose journal (ao, aop, josaa, josab, ol, pr)

\setboolean{shortarticle}{true}
\usepackage{amsthm,amsmath,amssymb}

\usepackage{mathrsfs}
\usepackage{color}

\title{Effects of Transceiver Jitter on the Performance of Optical Scattering Communication Systems}

\author[1,2]{Zanqiu Shen}
\author[1]{Jianshe Ma}
\author[3]{Serge B. Provost}
\author[1,*]{Ping Su}

\affil[1]{Division of Advanced manufacturing, Tsinghua Shenzhen International Graduate School, Shenzhen 518055, China}
\affil[2]{Department of Precision Instrument, Tsinghua University, Beijing 10084, China}
\affil[3]{Department of Statistical and Actuarial Sciences, The University of Western Ontario, London, Canada, N6A 5B7}

\affil[1,*]{Corresponding author: su.ping@mail.sz.tsinghua.edu.cn}

\dates{Published October 7, 2020}

\doi{\url{https://doi.org/10.1364/OL.400969}}

\begin{abstract}
In ultraviolet communications, the transceiver jitter effects have been ignored in previous studies, which can result in non-negligible performance degradation especially in vibration states or in mobile scenes. To address this issue, we model the relationship between the received power and transceiver jitter by making use of a moment-based density function approximation method. Based on this relationship, we incorporate the transceiver jitter effects in combination with Poisson distribution. The error rate results are obtained assuming on-off key modulation with optimal threshold based detection. We validate the error rate expressions by comparing the analytical results with Monte-Carlo simulation results. The results show that the transceiver jitter effects cause performance degradation especially in smaller transceiver elevation angles or in shorter distances, which are often adopted in short-range ultraviolet communications. The results also show that larger elevation angle cases have a better performance with respect to anti-jitter and may perform better compared to smaller elevation angle situations in the case of larger standard deviation of jitter. This work studies for the first time the transceiver jitter effects in ultraviolet communications and provides guidelines for experimental system design.
\end{abstract}

\setboolean{displaycopyright}{true}

\begin{document}

\maketitle

Optical scattering communication (OSC) has received increasing attention due to its applicability in non-line-of-sight (NLOS) communication scenarios \cite{vavoulas2019survey,yuan2016review}. There have been extensive studies on NLOS channel modeling including single scattering models in integration-form \cite{luettgen1991non,shen2019modeling}, approximation models in closed-form \cite{xu2008analytical,zuo2013closed}, Monte-Carlo (MC) models incorporating multiple scattering effects \cite{ding2009modeling,yuan2019monte}, and MC models with height difference \cite{song2017performance}. The theoretical analyses and experimental results both reveal that NLOS ultraviolet communication incurs large path loss, which results in weak received signal. Therefore, a photon counting receiver is adopted in the experiment and the experimental results show that the Poisson distribution is somewhat preferable for approximating photon counting distributions \cite{chen2008experimental}. However, all of the above studies assume constant transceiver geometric parameters, and neglect the mechanical jitter of OSC systems, which may result in random variations in the transceiver elevation angles and azimuth angles. Neglecting these mechanical jitter effects may underestimate the performance degradation, especially in vibration states or in mobile scenes, such as an unmanned aerial vehicle, a car in motion or a building sway. For example, if unmanned aerial vehicles are equipped with OSC systems, there are $\pm 0.02\,{\rm rad}$ vibrations in elevation angles and $\pm 0.1\,{\rm rad}$ in azimuth angles \cite{wu2020energy}.

In free space optical (FSO) communication, the pointing error is studied using the Rayleigh distribution in combination with turbulence effects \cite{arnon2003effects}, and  the channel capacity is evaluated using a wave-optics-based approach by considering the pointing fluctuations \cite{borah2009pointing}. In connection with underwater communication, Tang \emph{et al.} investigated the performance of link misalignment and suggested that an appropriate receiver offset distance might be permitted under certain conditions \cite{tang2012Onlink}. In the case of OSC systems, the drop of received optical power caused by misalignment can be significant \cite{arya2020novel}. For example, it has been revealed that when the transceiver elevation angles or the transceiver azimuth angles vary, the path loss will change \cite{shen2019modeling,zuo2013closed}, indicating that the jitter of transceiver parameters may cause non-negligible performance degradation. Therefore, it should prove worthwhile to study the effects of the transceiver jitter and provide guidelines for transceiver parameters setting. To the best of our knowledge, the transceiver jitter effects have not been studied in OSC systems.

In this letter, we propose a jitter model for characterizing the transceiver jitter effects. The turbulence effects are not considered, since it can be ignored when the communication range is less than $200\,{\rm m}$ \cite{liao2015uv}. We use a Maclaurin series to approximate the received power in both the coplanar and non-coplanar cases. Then the series forms are transformed into indefinite quadratic forms in normal vectors with shift $\epsilon$. The indefinite quadratic forms are then expressed as linear combinations of non-central chi-square random variables. A moment-based density function approximation method is adopted to derive the probability density function (PDF) of the received power, which takes into account the jitter effects. Then we incorporate the jitter effects using a Poisson distribution and obtain the probability mass function (PMF) of the number of the received photons. Finally, we derive the error probability for a photon counting receiver using on-off key (OOK) modulation with optimal threshold based detection. Compared with the model introduced in \cite{zuo2013closed}, the proposed model not only considers transceiver configurations, atmospheric scattering and absorbing effects, but also takes into account the jitter effects of the transceiver elevation and azimuth angles. Next, we present a detailed derivation of the proposed model.

In the case of ultraviolet communications, the received signal is weak and the received photons are assumed to follow a Poisson distribution. We define $\lambda_s$ as the mean number of photons of the received signal and $\lambda_b$ as the mean number of photons of the background radiation noise. Under this definition, when the transmitter transmits symbol "1", the number of received photons can be modeled as 
\begin{equation}
    \mathbb{P}(n; \lambda_s+\lambda_b) = \frac{(\lambda_s+\lambda_b)^n}{n!} \exp[-(\lambda_s+\lambda_b)].
\label{symbol "1":possion distribution}
\end{equation}
The photon counting rate is given by$ \Lambda = {\eta E_r \lambda}/({h c}),\label{possion counting rate}$ where $\eta$ is the quantum efficiency of the detector including optical filter $\eta_f$ and photodetector $\eta_{p}$, $E_r$ is the received power, $h$ is the Planck constant, $c$ is the optical speed and $\lambda$ is the optical wavelength. The mean number of photos $\lambda_s$ of the received signal can be obtained as $\Lambda T_p$, where $T_p$ is the duration time of one pulse. The mean number of photons $\lambda_b$ of the background radiation noise can be obtained as $N_n T_p$, where $N_n$ is the background noise counting rate with typical values of $0, 5{\rm kHz}$ and $14.5 {\rm kHz}$ based on measurements \cite{chen2008experimental}.

When the transmitter transmits symbol "0", the received number of photons is only dominated by the background radiation noise and it can then be expressed as
\begin{equation}
    f(n; \lambda_b) = \frac{\lambda_b^n}{n!} \exp[-\lambda_b].
\label{symbol "0":possion distribution}
\end{equation}
Eq. (\ref{symbol "1":possion distribution}) and (\ref{symbol "0":possion distribution}) assume a constant transceiver geometric configuration, which may underestimate the performance degradation, especially when the transceiver works in high speed vibration states. We define $f{(E_r)}$ as the probability density function of the received power caused by random variation of the transceiver geometric parameters. Combined with Eq. (\ref{symbol "1":possion distribution}), the received number of photons $n$ satisfies
\begin{equation}
\begin{split}
f(n;\lambda_s+\lambda_b) = \int_{0}^{+\infty} \frac{(\lambda_s(E_r)+\lambda_b)^n}{n!} \exp[-(\lambda_s(E_r)+\lambda_b)] \\
\cdot f(E_r) {\rm d} E_r.
\label{Received photon numbers consider geometric change}
\end{split}
\end{equation}
In this case, the transmitter beam angles and receiver field of views (FOV) are constant while elevation angles and azimuth angles change randomly. To characterize the random effects of the geometric parameters, we should use a non-coplanar path-loss model. An accurate and closed-form non-coplanar path loss model has been proposed in \cite{zuo2013closed}. Following \cite{zuo2013closed}, we obtain the received power $E_r$ as follows:
\begin{equation}
E_r = \frac{E_t A_r {\alpha_T}^2 p(\cos{\theta_s}) \cos{\zeta}}{4 \Omega_t k_e r^{'2} \exp(k_e r^{'})} [\exp(-k_e r_A) - \exp(-k_e r_B)],
\label{Received energy Er}
\end{equation}
where $E_t$ is the transmission power in one pulse, $A_r$ is the receiver area, $\alpha_T$ is the half beam angle of the transmitter, $\theta_s$ is the angle between the transceiver axes, $p(\cos(\theta_s))$ is the scattering phase function as a weighted sum of Rayleigh and Mie type scattering phase functions \cite{xu2008analytical}, $\zeta$ is the angle between the scattered direction and receiver FOV axis, $\Omega_t=2\pi(1-\cos{(\alpha_T)})$ is the transmitter solid cone angle, $k_e$ is the extinction coefficient as a sum of Rayleigh scattering coefficient $k_r$, Mie scattering coefficient $k_m$ and atmospheric absorption coefficient $k_a$, $r_A$ and $r_B$ are the distances from the transmitter to A and B as illustrated in \cite{zuo2013closed}, and $r^{'}$ is the distance from $\delta V$ to the receiver as defined in \cite{zuo2013closed}. 

Considering the jitter effects of the transceiver elevation angles and azimuth angles, we replace $(\theta_T, \theta_R)$ and $(\phi_T, \phi_R)$ with $(\theta_T+\theta_{Tj},\theta_R+\theta_{Rj})$ and $(\phi_{T}+\phi_{Tj}, \phi_R+\phi_{Rj})$, where ($\theta_{Tj}, \theta_{Rj}, \phi_{Tj}, \phi_{Rj}$) are random variables denoting the jitter effects of transceiver elevation angles and azimuth angles. The jitter index parameters are the means $(\mu_{\theta_{Tj}}, \mu_{\phi_{Tj}}, \mu_{\theta_{Rj}}, \mu_{\phi_{Rj}})$ and the standard deviations $(\sigma_{\theta_{Tj}}, \sigma_{\theta_{Rj}}, \sigma_{\phi_{Tj}}, \sigma_{\phi_{Rj}})$ of the random variables ($\theta_{Tj}, \theta_{Rj}, \phi_{Tj}, \phi_{Rj}$). Then the received power $E_r$ can be written as a function of random variables i.e., $f_{E_r}(\theta_{Tj}, \theta_{Rj}, \phi_{Tj}, \phi_{Rj})$. Since $(\theta_{Tj}, \theta_{Rj}, \phi_{Tj}, \phi_{Rj})$ vary around zero, we use Maclaurin series to expand $f_{E_r}(\theta_{Tj}, \theta_{Rj}, \phi_{Tj}, \phi_{Rj})$. For tractable analysis, $f_{E_r}(\theta_{Tj}, \theta_{Rj}, \phi_{Tj}, \phi_{Rj})$ is approximated by first-order and second-order partial derivatives,  which are
\begin{subequations}
\begin{equation}
\begin{split}
f_{E_r}(\theta_{Tj}, \theta_{Rj}, \phi_{Tj}, \phi_{Rj}) \approx f_{E_r}(0,0,0,0)+f_1 \theta_{Tj}+\\
f_2 \theta_{Rj}
+f_3 \phi_{Tj}+f_4 \phi_{Rj}+\alpha F \alpha^T,
\end{split}
\end{equation}
where 
\begin{equation}
  (f_1, f_2, f_3, f_4) = {\Big (}\frac{\partial{f_{E_r}}}{\theta_{Tj}}, \frac{\partial{f_{E_r}}}{\theta_{Rj}}, \frac{\partial{f_{E_r}}}{\phi_{Tj}}, \frac{\partial{f_{E_r}}}{\phi_{Rj}}{\Big)},
\end{equation}
\begin{equation}
   \alpha = (\theta_{Tj}, \theta_{Rj}, \phi_{Tj}, \phi_{Rj}),
\end{equation}
$F=$
\begin{equation}
\begin{split}
\setlength{\abovedisplayskip}{3pt}
\left[
\begin{matrix}
	\frac{\partial^2{f_{E_r}}}{\partial \theta_{Tj}^2}& \frac{\partial^2{f_{E_r}}}{\partial \theta_{Tj} \partial \theta_{Rj}}& \frac{\partial^2{f_{E_r}}}{\partial \theta_{Tj} \partial \phi_{Tj}}& \frac{\partial^2{f_{E_r}}}{\partial \theta_{Tj} \partial \phi_{Rj}} \\
	\frac{\partial^2{f_{E_r}}}{\partial \theta_{Rj} \partial \theta_{Tj}}& \frac{\partial^2{f_{E_r}}}{\partial \theta_{Rj} \partial \theta_{Rj}}& \frac{\partial^2{f_{E_r}}}{\partial \theta_{Rj} \partial \phi_{Tj}}& \frac{\partial^2{f_{E_r}}}{\partial \theta_{Rj} \partial \phi_{Rj}} \\
	\frac{\partial^2{f_{E_r}}}{\partial \phi_{Tj} \partial \theta_{Tj}}& \frac{\partial^2{f_{E_r}}}{\partial \phi_{Tj} \partial \theta_{Rj}} & \frac{\partial^2{f_{E_r}}}{\partial \phi_{Tj} \partial \phi_{Tj}}& \frac{\partial^2{f_{E_r}}}{\partial \phi_{Tj} \partial \phi_{Rj}} \\
    \frac{\partial^2{f_{E_r}}}{\partial \phi_{Rj} \partial \theta_{Tj}}& \frac{\partial^2{f_{E_r}}}{\partial \phi_{Rj} \partial \theta_{Rj}} & \frac{\partial^2{f_{E_r}}}{\partial \phi_{Rj} \partial \phi_{Tj}}& \frac{\partial^2{f_{E_r}}}{\partial \phi_{Rj} \partial \phi_{Rj}}  
\end{matrix}
\right]
\equiv \left[
\begin{matrix}
	f_{11}& f_{12}& f_{13}& f_{14}\\
	f_{21}& f_{22}& f_{23}& f_{24}\\
	f_{31}& f_{32}& f_{33}& f_{34}\\
	f_{41}& f_{42}& f_{43}& f_{44}
\end{matrix}
\right].
\end{split}
\setlength{\abovedisplayskip}{3pt}
\end{equation}
\label{Taylor approx}
\end{subequations}
The partial derivatives can easily be computed by {\sl Mathematica} 12.1 and so, are omitted. To continue the derivation, we express Eq. (\ref{Taylor approx}) in matrix form with shift $\epsilon$ as follows:
\begin{subequations}
\setlength{\abovedisplayskip}{3pt}
\begin{equation}
\setlength{\abovedisplayskip}{3pt}
\begin{split}
f_{E_r}(\theta_{Tj}, \theta_{Rj}, \phi_{Tj}, \phi_{Rj}) \approx \beta G \beta^T+\epsilon,
\end{split}
\setlength{\abovedisplayskip}{3pt}
\end{equation}
where
\begin{equation}
\setlength{\abovedisplayskip}{3pt}
    \epsilon =  f_{E_r}(0,0,0,0)+e,
\setlength{\abovedisplayskip}{3pt}
\end{equation}
\begin{equation}
\setlength{\abovedisplayskip}{3pt}
\begin{split}
   \beta& = (\theta_{Tj}-a, \theta_{Rj}-b, \phi_{Tj}-c, \phi_{Rj}-d)\\
        & = (\beta_1, \beta_2, \beta_3, \beta_4).
\end{split}
\setlength{\abovedisplayskip}{3pt}
\end{equation}
\label{Matrix form}
\setlength{\abovedisplayskip}{3pt}
\end{subequations}
By solving the equation group, we observe that matrix $G$ is the same as matrix $F$. The parameters $a, b, c, d$ can be determined by Cramer's Rule. Defining $(G_i|i=1,2,3,4)$ as the matrix by substituting the $i_{th}$ column of $G$ with the column vector $(-f_1/2, -f_2/2, -f_3/2, -f_4/2)^{\rm T}$, we obtain $a = \frac{{\rm det}(G_1)}{{\rm det}(G)}$, $b = \frac{{\rm det}(G_2)}{{\rm det}(G)}$, $c = \frac{{\rm det}(G_3)}{{\rm det}(G)}$, $d = \frac{{\rm det}(G_4)}{{\rm det}(G)}$, and
\begin{equation}
\begin{split}
e = -(a^2 f_{11}+2 a b f_{12}+2 a c f_{13}+2 a d f_{14} + b^2 f_{22}+2 b c f_{23}+\\
2 b d f_{24} + c^2 f_{33} + 2 c d f_{34} + d^2 f_{44}),
\end{split}
\end{equation}
where ${\rm det}({\rm A})$ denotes the determinant of the matrix $A$. We model the jitter effects of transceiver elevation angles and azimuth angles with Gaussian distribution, satisfying $\beta_1 \sim \mathcal{N}(-a, \sigma_{\theta_{Tj}}^2)$, $\beta_2 \sim \mathcal{N}(-b, \sigma_{\theta_{Rj}}^2)$,
$\beta_3 \sim \mathcal{N}(-c, \sigma_{\phi_{Tj}}^2)$,
and $\beta_4 \sim \mathcal{N}(-d, \sigma_{\phi_{Rj}}^2)$. In this case, we can see that the received power is a quadratic form in normal vectors with a shift $\epsilon$, which can be transformed into linear combinations of non-central chi-square random variables with a shift $\epsilon$. Combined with the shift $\epsilon$, the probability density function of $E_r$ can be determined as follows \cite{provost2005moment}:
\begin{subequations}
\setlength{\abovedisplayskip}{3pt}
\begin{equation}
\setlength{\abovedisplayskip}{3pt}
	f(E_r) = \left\{
	\begin{array}{rcl}
	h_{1}(E_r-\epsilon) &   &{if\, (E_r-\epsilon) > 0}\\
	h_{2}(E_r-\epsilon) &   &{if\, (E_r-\epsilon) \leq 0}
	\end{array} \right.
\setlength{\abovedisplayskip}{3pt}
\end{equation}
where 
\begin{equation}
\setlength{\abovedisplayskip}{3pt}
g_1(x) = \frac{\exp(-\frac{x}{\beta_1})}{\Gamma(\alpha_1) \beta_{1}^{\alpha_1}} \sum_{i=0}^{q} m_i {x}^{i+\alpha_1-1},
\end{equation}
\begin{equation}
\setlength{\abovedisplayskip}{3pt}
g_2(x) = \frac{\exp(-\frac{x}{\beta_2})}{\Gamma(\alpha_2) \beta_{2}^{\alpha_2}} \sum_{j=0}^{q} n_j {x}^{j+\alpha_2-1},
\end{equation}
\begin{equation}
\setlength{\abovedisplayskip}{3pt}
h_{1}(x) = \int_{0}^{\infty} g_{1}(x+y)g_{2}(y) {\rm d}y,
\end{equation}
\begin{equation}
\setlength{\abovedisplayskip}{3pt}
h_{2}(x) = \int_{-x}^{\infty} g_{1}(x+y)g_{2}(y) {\rm d}y.
\end{equation}
\label{f(E_r)}
\setlength{\abovedisplayskip}{3pt}
\end{subequations}
The appropriate value of $q$ can be chosen to achieve a desired accuracy. The detailed derivation of $\alpha_1, \beta_1, \alpha_2, \beta_2, (m_i;i=0,1,...,q)$ and $(n_j;j=0,1,...,q)$ can be found in \cite{provost2005moment,ha2013accurate}. Substituting Eq. (\ref{f(E_r)}) into Eq. (\ref{Received photon numbers consider geometric change}), we obtain the PMF of the received number of photons as follows:
\begin{equation}
\setlength{\abovedisplayskip}{2pt}
\begin{split}
f(n;\lambda_s+\lambda_b) = \int_{0}^{\epsilon} \frac{(\lambda_s(E_r)+\lambda_b)^n}{n!} \exp[-(\lambda_s(E_r)+\lambda_b)] \\
\cdot \int_{-(E_r-\epsilon)}^{\infty} g_1(E_r-\epsilon+y)g_2(y){\rm d}y {\rm d}E_r+\int_{\epsilon}^{\infty} \frac{(\lambda_s(E_r)+\lambda_b)^n}{n!}\\ \cdot \exp[-(\lambda_s(E_r)+\lambda_b)]\int_{0}^{\infty} g_1(E_r-\epsilon+y)g_2(y){\rm d}y {\rm d}E_r.
\label{PMF of the received number of photons}
\end{split}
\setlength{\abovedisplayskip}{2pt}
\end{equation}
The corresponding cumulative distribution function (CDF) is then
\begin{equation}
\setlength{\abovedisplayskip}{2pt}
\begin{split}
F(k;\lambda_s+\lambda_b) = \sum_{n=0}^{k} f(n;\lambda_s+\lambda_b).
\label{CDF of the received number of photons}
\end{split}
\setlength{\abovedisplayskip}{3pt}
\end{equation}

Next, we consider the case of a photon counting receiver with optimal threshold based OOK modulation whose error probability can be obtained as $ P_{e} = P_1 \sum_{n=0}^{n_{th}} f(n;\lambda_s+\lambda_b) + P_0 \sum_{n=n_{th}+1}^{+\infty} f(n;\lambda_b),$ where $n_{th}$ is the threshold which can be optimized by minimizing $P_{e}$ and $P_1=P_0=0.5$ is assumed in this work. When the transceiver jitter effects are ignored, the error probability is given in \cite{he2010performance}. When the transceiver jitter effects are taken into account, the error probability can be obtained as
\begin{equation}
\setlength{\abovedisplayskip}{2pt}
\begin{split}
P_{ej} = \frac{1}{2} \sum_{n=0}^{n_{th}} \Big[ \int_{0}^{\epsilon} \frac{(\lambda_s(E_r)+\lambda_b)^n}{n!} \exp[-(\lambda_s(E_r)+\lambda_b)] \\
\cdot \int_{-(E_r-\epsilon)}^{\infty} g_1(E_r-\epsilon+y)g_2(y){\rm d}y {\rm d}E_r+\int_{\epsilon}^{\infty} \frac{(\lambda_s(E_r)+\lambda_b)^n}{n!}\\ \cdot \exp[-(\lambda_s(E_r)+\lambda_b)]\int_{0}^{\infty} g_1(E_r-\epsilon+y)g_2(y){\rm d}y {\rm d}E_r \Big]  \\
+ \frac{1}{2} \sum_{n=n_{th}+1}^{\infty} \frac{\lambda_b^n}{n!} \exp[-\lambda_b],
\label{P_e: jitter}
\end{split}
\setlength{\abovedisplayskip}{2pt}
\end{equation}
where $n_{th}$ can be determined by searching from $floor(\lambda_b)$ to $floor(\lambda_s)$ to minimize $P_{ej}$. Next, we present our numerical results.

\begin{figure}[ht]
\setlength{\abovecaptionskip}{0cm}
\setlength{\belowcaptionskip}{-0.5cm}
\centering
\includegraphics[width=\linewidth]{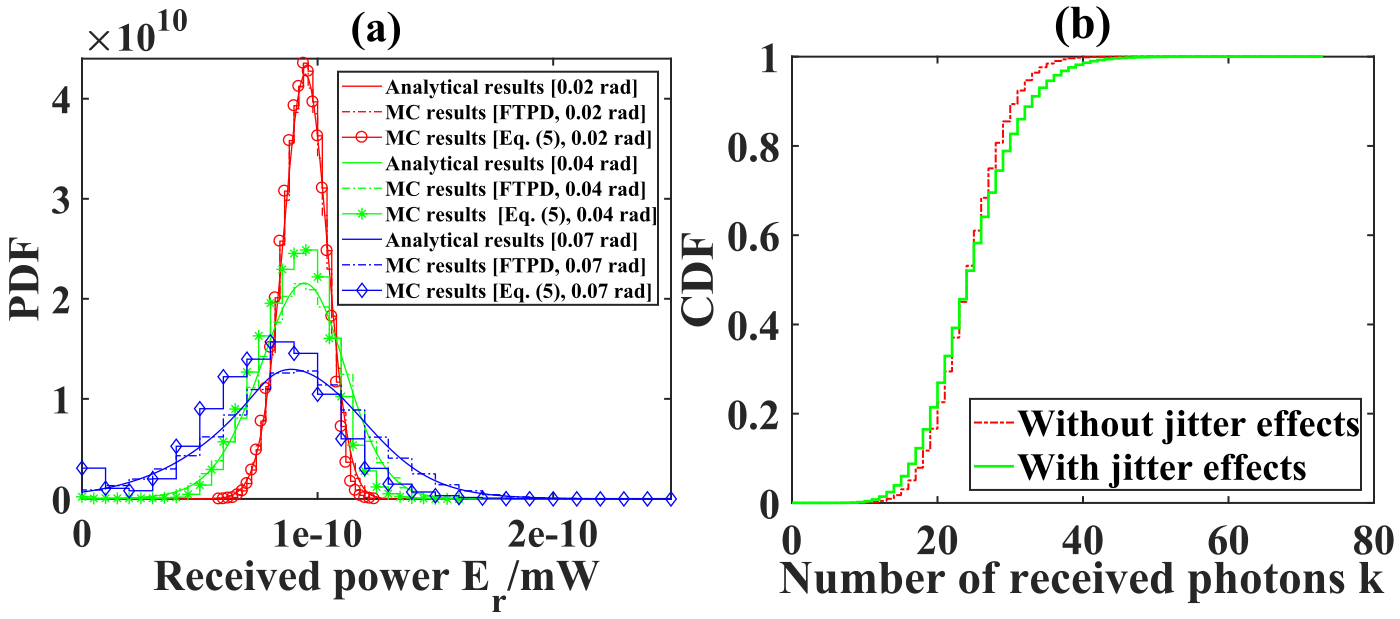}
\caption{Geometric setting: $(\theta_T, \theta_R, \phi_T, \phi_R)=(20^{\circ}, 20^{\circ}, 5^{\circ}, 0^{\circ})$. (a). The analytical results and MC simulation results of the PDF of received power; (b). The CDFs with and without transceiver jitter effects.}
\label{fig:1}
\end{figure}

In Fig. \ref{fig:1}(a), we validate the analytical PDF in Eq. (\ref{f(E_r)}) by MC simulations of Eq.(\ref{Received energy Er}) and Eq. (\ref{Matrix form}). we calculate the received power by generating $10^5$ samples and obtain the corresponding frequency histograms in MC simulations. The simulation parameters are set as follows: $A_r=1.77\,{{\rm cm}^2}$, $(k_r, k_m, k_a)=(0.266, 0.284, 0.802)\,{{\rm km}^{-1}}$ at $\lambda=260\,{\rm nm}$, $(\gamma, g, f)=(0.017, 0.72, 0.5)$, $N_n = 14500\,{\rm s}^{-1}$, $(\eta_f,\eta_p)=(0.2, 0.3)$, $r=50\,{\rm m}$, parameter $q=6$, half beam and half FOV $(\alpha_T,\alpha_R)=(1^{\circ}, 30^{\circ})$, and average transmission power $P_t=50\,{\rm mW}$. It is noted that the scattering coefficients and absorption coefficient are chosen on the basis of wavelength \cite{ding2009modeling}. The transmission power $E_t$ in one pulse is given as $2P_t$ for OOK modulation. We note that different means correspond to different non-coplanar transceiver configurations, and a model reflecting this has been established and analyzed in a previous study \cite{zuo2013closed}. Therefore, we focus on the analysis of the effects of the standard deviations of ($\theta_{Tj}, \theta_{Rj}, \phi_{Tj}, \phi_{Rj}$). The means of ($\theta_{Tj}, \theta_{Rj}, \phi_{Tj}, \phi_{Rj}$) are given as $(\mu_{\theta_{Tj}}, \mu_{\phi_{Tj}}, \mu_{\theta_{Rj}}, \mu_{\phi_{Rj}})$ = $(0^{\circ},0^{\circ},0^{\circ},0^{\circ})$. In Fig. \ref{fig:1}(a), three cases were simulated with different jitter standard deviations, which are 0.02 rad, 0.04 rad, and 0.07 rad. For each case, we provided the simulation results corresponding to analytical expressions, the MC simulation using Eq. (\ref{Matrix form}) based on first two partial derivative (FTPD) approximation, and the MC simulation using Eq. (\ref{Received energy Er}). It is seen from Fig. \ref{fig:1}(a) that the analytical results closely agree with the MC simulation results using Eq. (\ref{Matrix form})based on FTPD approximation for each case, which indicates that $q = 6$ is adequate in connection with analytical results. Furthermore, we can see that the MC simulation results using Eq. (\ref{Received energy Er}) are also in close agreement with the analytical results obtained when the jitter standard deviations are small. However, the deviations between the MC simulation results using Eq. (\ref{Matrix form}) and the analytical results increase when the jitter standard deviations increase, which means that the influence of the third and higher order partial derivatives becomes more significant as the jitter standard deviations increase. Although the FTPD approximation results may not be as accurate in the case of large standard deviations, they can still be used to derive the probability density function of the received power. In Fig. \ref{fig:1}(b), the standard deviation is set to $0.04\,{\rm rad}$ and it can be seen that the CDF with transceiver jitter effects is flatter than the CDF without jitter effects, which indicates that the variance increases when the transceiver jitter effects are combined.

\begin{figure}[ht]
\setlength{\abovecaptionskip}{0cm}
\setlength{\belowcaptionskip}{-0.5cm}
\centering
\includegraphics[width=\linewidth]{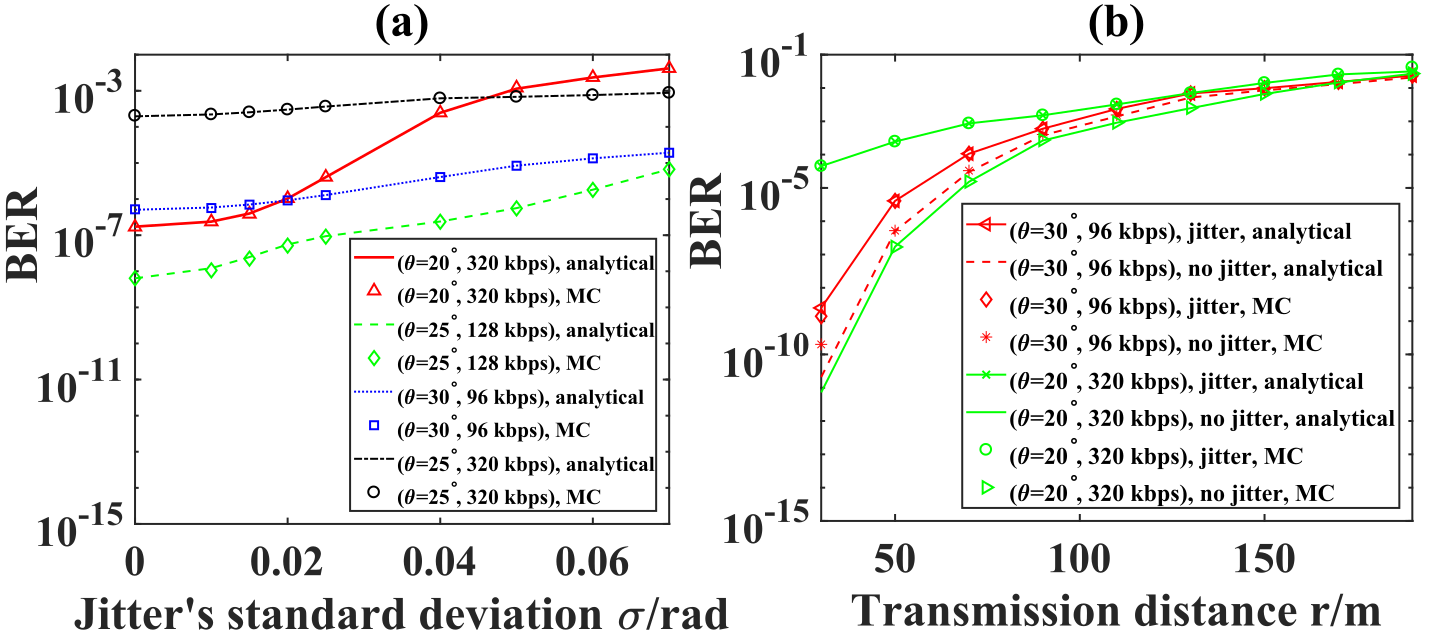}
\caption{Geometric setting: $(\phi_T, \phi_R)=(5^{\circ}, 0^{\circ})$. (a). BERs in different jitter's standard deviations with different transceiver configurations and typical data rates; (b). BERs in different transmission distances with different transceiver configurations and typical data rates.}
\label{fig:2}
\end{figure}

In Fig. \ref{fig:2}, we plot the bit error rates (BERs) versus the standard deviations $\sigma$ of jitter and the transmission distances in different transceiver configurations with typical data rates according to Eq. (\ref{P_e: jitter}). We assume $\theta_T=\theta_R=\theta$ for simplicity. To validate the analytical results, we simulated $5\times10^9$ symbols with the MC simulation method using the CDF in Eq. (\ref{CDF of the received number of photons}). From Fig. \ref{fig:2}(a), we see that the analytical results are in agreement with the MC simulation results. Moreover, the performance degradation is observed from Fig. \ref{fig:2}(a) for four transceiver configurations; they are especially noticeable when the transceiver elevation angles are relatively small. Smaller transceiver elevation angle cases are often preferred to achieve higher data rates since these cases have smaller path loss. 
However, it is seen from Fig. \ref{fig:2}(a) that the smaller elevation angle cases are more sensitive to transceiver jitter compared to larger transceiver elevation angle cases. For example, when the standard deviation $\sigma$ changes from 0 to 0.07 rad, the BERs increase from $1.7\times 10^{-7}$ to $4.2\times10^{-3}$ for $(\theta_T=\theta_R=\theta=20^{\circ}, 320{\rm kbps})$ while the BERs increase from $5.1\times10^{-7}$ to $1.9\times10^{-5}$ for $(\theta_T=\theta_R=30^{\circ}, 96{\rm kbps})$. Therefore, when the elevation angles are smaller, the jitter needs to be reduced to achieve a desired error rate. Besides, we can see that when the standard deviations $\sigma$ are smaller than 0.046 rad, the BERs for $(\theta_T=\theta_R=\theta=20^{\circ}, 320{\rm kbps})$ are smaller than the BERs for $(\theta_T=\theta_R=\theta=25^{\circ}, 320{\rm kbps})$. However, when the standard deviations $\sigma$ are greater than 0.046 rad, the BERs for $(\theta_T=\theta_R=\theta=20^{\circ}, 320{\rm kbps})$ are greater than the BERs for $(\theta_T=\theta_R=\theta=25^{\circ}, 320{\rm kbps})$. This suggests that although larger elevation angle cases have larger path loss, these cases have a better anti-jitter performance than smaller elevation angle cases. In Fig. \ref{fig:2}(b), we set the standard deviation to $\sigma=0.04\,{\rm rad}$ as an example. It is seen from Fig. \ref{fig:2}(b) that the analytical results are in close agreement with the MC simulation results, and the deviations between these two results in smaller distance cases are due to the limitation of the computing resource. Moreover, the BERs with jitter effects are larger than those observed without jitter effects, and the influence of jitter is more significant when the transmission distances are limited. Taking ($\theta_T=\theta_R=\theta=30^{\circ}, 96\,{\rm kbps}$) as an example, when the transmission distance is 30 m, the BER with jitter effects, which is $2.5\times\,10^{-9}$, is two orders of magnitude higher than the BER observed without jitter effects, which is $2.1\times\,10^{-11}$. However, when the transmission distance is 170 m, the BERs evaluated with and without jitter effects are almost the same.

In summary, we modeled the relationship between the received power and the transceiver jitter effects with Maclaurin series. Based on this relationship, we incorporated the transceiver jitter effects with Poisson distribution. Moreover, we obtained the error probability of OOK modulation assuming optimal threshold based detection. The results show that the transceiver jitter causes performance degradation in different transceiver geometries, and smaller transceiver elevation angle cases and shorter distance cases are more sensitive to transceiver jitter. The results also show that larger elevation angle cases may have a better performance compared to smaller elevation angle cases in larger standard deviations. In future work, we will focus on performance improvement technologies to counter transceiver jitter effects.
\\ \\
{\bfseries Funding.} This work was supported by the Basic Research Program of Shenzhen (JCYJ20170412171744267).

\medskip
\noindent\textbf{Disclosures.} The authors declare that there are no conflicts of interest related to this article.
\medskip
% Bibliography
\bibliography{sample}
\bibliographyfullrefs{sample}

\end{document}